\documentclass[10pt,aps,prl, twocolumn,showpacs,superscriptaddress,longbibliography]{revtex4-2}
\usepackage{xcolor}
\usepackage{graphicx}
\usepackage{bm}
\usepackage{enumitem}
\usepackage{amsmath}
\usepackage{bbold}
\usepackage{mathtools}
\usepackage{amsmath, bigstrut}
\usepackage{amssymb}
\usepackage{mathrsfs}
\usepackage{notoccite}
\usepackage{pstricks}
\usepackage[normalem]{ulem}
\usepackage{float}
\usepackage{url}
\usepackage{hyperref}

\makeatletter
\renewcommand{\fnum@figure}{\textbf{Figure \thefigure}}
\makeatother

\def\papertitle{
	Structured detection microscopy
}

\begin{document}

\title{\bfseries \boldmath \papertitle}

\author{Larnii Booth}
\thanks{These authors contributed equally.}
\affiliation{School of Mathematics and Physics, The University of Queensland, Brisbane, Australia}

\author{Kyle Clunies-Ross}
\thanks{These authors contributed equally.}
\affiliation{School of Mathematics and Physics, The University of Queensland, Brisbane, Australia}
\affiliation{ARC Centre of Excellence in Quantum Biotechnology, Brisbane, Australia}

\author{Rumelo Amor}
\affiliation{Queensland Brain Institute, The University of Queensland, Brisbane, Australia}

\author{Nicolas Mauranyapin}
\affiliation{School of Mathematics and Physics, The University of Queensland, Brisbane, Australia}
\affiliation{ARC Centre of Excellence in Quantum Biotechnology, Brisbane, Australia}

\author{Zixin Huang}
\affiliation{School of Science, College of STEM, RMIT University, Melbourne, Australia}

\author{Michael A. Taylor}
\affiliation{Department of Anatomy and Department of Physics, University of Otago, Dunedin, New Zealand}

\author{Warwick P. Bowen}
\thanks{Corresponding author: w.bowen@uq.edu.au}
\affiliation{School of Mathematics and Physics, The University of Queensland, Brisbane, Australia}
\affiliation{ARC Centre of Excellence in Quantum Biotechnology, Brisbane, Australia}
\email{w.bowen@uq.edu.au}



\begin{abstract} \bfseries \boldmath
\noindent Super-resolution microscopy is crucial for imaging sub-wavelength biological structures. However, most techniques rely on nonlinear saturation or stochastic switching of emitters, limiting imaging speed and increasing phototoxicity. Here, we achieve deep super-resolution without employing saturation or stochastic dynamics, instead using a form of spatial mode demultiplexing. By shaping the point-spread function of the emitted light, our Structured Detection Microscope (SDM) redistributes information away from high shot-noise regions of the image, enhancing sensitivity to sub-diffraction emitter separations in two-dimensions and without mode-sorting optics. Implementing SDM within a high-numerical aperture total internal reflection fluorescence microscope, we demonstrate imaging of fluorophores attached to DNA nanorulers with separations as small as 50~nm at resolutions surpassing 40~nm -- fivefold below the diffraction limit. This shows that spatial mode demultiplexing can achieve far sub-wavelength resolution and is applicable to biologically relevant samples. By enabling super-resolution biomolecular imaging without emitter saturation and stochasticity, our work opens the door to better understanding biological structure, function and dynamics.
\end{abstract}

\maketitle

\clearpage

\section*{Introduction}
Optical microscopy is indispensable in the life sciences~\cite{Hell:2015}, yet its resolution is generally constrained by the diffraction limit, first described by Abbe~\cite{Abbe:1873}. Many fundamental biological structures are substantially smaller than this limit~\cite{Hell:2007,Mauranyapin:2022,Taylor:2016}, motivating research on super-resolution approaches~\cite{Prakash:2025}. Over the past three decades, techniques such as stimulated emission depletion (STED) \cite{Hell:94} and single-molecule localization~\cite{Betzig:2006,Rust:2006,Heilemann:2008,Sharonov:2006,Schnitzbauer:17,Balzarotti:17} have revolutionized fluorescence microscopy, achieving molecular- to angstrom-level resolution~\cite{Balzarotti:17}. However, their dependence on nonlinear saturation or stochastic dynamics introduces complexity, prolongs acquisition times, and constrains applications in dynamic or high-throughput live cell studies~\cite{Prakash:2024}. Structured Illumination Microscopy (SIM) overcomes some of these issues, but without photoswitchable fluorescence techniques~\cite{Rego:2012}, its resolution is capped at $\sim$100 nm \cite{Ortkrass:23}. 

It has recently been shown that the diffraction limit can be circumvented using spatial mode demultiplexing (SPADE) of the emitted field~\cite{Tsang:16}. SPADE does not require nonlinearity or stochasticity and, in principle, can achieve deep sub-diffraction limited resolution~\cite{Tsang:16}. However, the best resolution reported to date is 220~nm, achieved in a proof-of-principle experiment on highly-scattering gold nanoparticles~\cite{Greenwood:2023}. Resolution can also be improved by engineering the point-spread function (PSF) of the emitted field~\cite{Paur:18}. This can be viewed as a form of SPADE where the demultiplexing is achieved by reconfiguring which spatial modes of the emission contribute to the intensity at each point in the image plane, and offers the distinct advantage of compatibility with standard cameras used for microscopy. However, experiments have so far been limited to emitter separations above 30~$\mu$m~\cite{Paur:18}. To date, neither conventional implementations of SPADE or point-spread function engineering has achieved far sub-wavelength resolution. To our knowledge, they have also not yet been implemented to image separations in two-dimensions. This has prevented applications in biomolecular imaging.

Here, we develop a two-dimensional PSF-engineered super-resolution microscope, which we call Structured Detection Microscopy (SDM) in analogy to SIM. By combining a phase plate to engineer the PSF in two dimensions, detection on a camera, and Bayesian analysis to optimally extract signal information, SDM allows super-resolved estimates of the separation and orientation of emitter pairs. We incorporate SDM into a state-of-the-art high-NA total internal reflection fluorescence (TIRF) microscope to enable far sub-wavelength imaging of biologically relevant specimens. We prove this capability by imaging pairs of fluourescent emitters attached to the ends of DNA nanorulers, showing that it is possible to resolve separations as small as 50~nm. Our modeling predicts that resolutions even beyond 5~nm could be achieved by increasing illumination intensity or acquisition time, while simultaneously reducing background counts. Broadly, our results show that spatial mode engineering can be applied to far sub-wavelength imaging in a biological microscope. This provides a tool to better understand biological structure, function and dynamics. 

\begin{figure*}
    \centering
    \includegraphics[width=\textwidth]{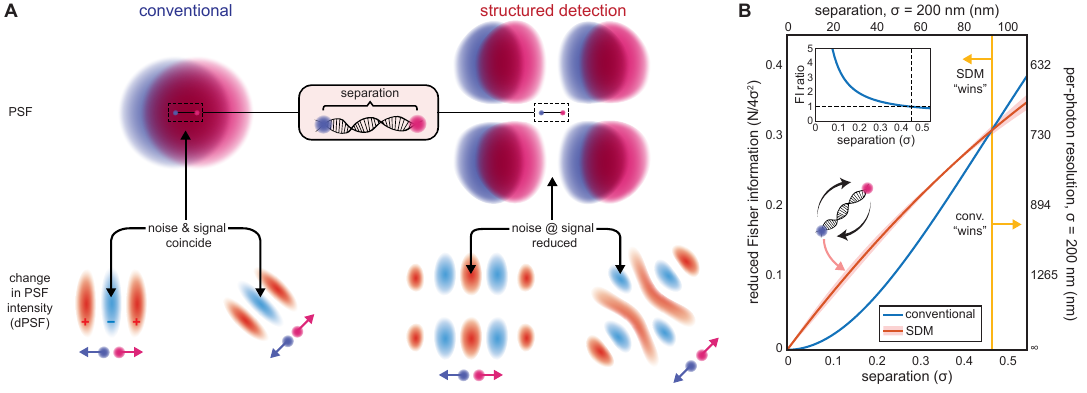}
    \caption{\textbf{Theoretical advantage of higher-order light modes in imaging.} (\textbf{A}) small shifts of the emitters from the centroid below the diffraction limit (indicated by the arrows on the emitters in the dPSF row) causes a signal change (indicated by the red/blue regions) in the point-spread function (PSF). The top-left diagram is the conventional microscopy signal, whose signal change is largest where shot noise (which is proportional to PSF intensity) is largest; the top-right diagram is the structured detection signal, whose change is largest where shot noise is smaller. (\textbf{B}) the effect on the Fisher information: for separations below a critical point of approximately 0.45$\sigma$, (with $\sigma$ the diffraction limit), SDM has a larger Fisher information (FI) and hence `wins' in imaging precision over conventional microscopy (vice versa for above 0.45$\sigma$). The inset shows the ratio of FI for SDM to conventional microscopy. There is a slight variance in the FI depending on emitter orientation for SDM, explained by the lack of circular symmetry in the structured detection PSF, however its impact is negligible compared to the gain in FI and makes structured detection effectively emitter-orientation independent.}
    \label{fig:theory}
\end{figure*}

\section*{Results}
\subsection*{Conceptual Overview}

\begin{figure*}
    \centering
    \begin{minipage}[hbt]{105mm}
    \vspace{0pt}
    \centering
    \includegraphics[width=\linewidth]{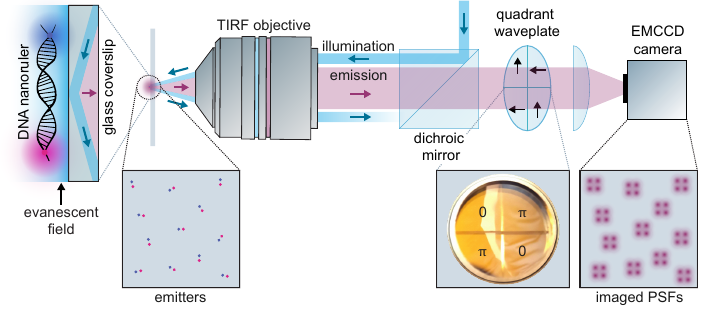}
    \end{minipage}
    \hfill
    \begin{minipage}[hbt]{70mm}
    \vspace{0pt}
    \caption{\textbf{Schematic of SDM experiment.} The illumination beam totally internally reflects off a glass coverslip with many DNA nanorulers affixed. The evanescent field excites the fluorophores attached to the ends of the nanorulers. A long-pass dichroic mirror then reflects the illumination beam and transmits the fluorescence emission. In the Fourier plane, the emission beam is shaped by a quadrant waveplate which has a $\pi$ phase shift on diagonally opposite quadrants (indicated by the arrows). In the image plane, the imaged PSFs show the expected mode shape. The quadrant waveplate can be removed to allow a comparison with conventional microscopy.}
    \label{fig:schematic}
    \end{minipage}
\end{figure*}

SDM improves spatial resolution by manipulating the PSF of the field to be imaged, building on the approach developed in~\cite{Paur:18,Paur:2019}. It does this by reshaping the PSF from the approximate Gaussian imaged in conventional microscopy. Here, our goal is to estimate the 2D spatial separation $\mathbf{s}$ between two closely spaced incoherent emitters, for which we find that reshaping the mode into a `split-Gaussian' PSF -- approximating a TEM$_{1,1}$ Hermite-Gauss mode -- is effective. 

The ultimate precision of any unbiased measurement is dictated by the Cram\'er–Rao bound (CRB)~\cite{Rao:1945,Cramer:1946}. In conventional imaging, the Fisher information decreases rapidly as $s=|\mathbf{s}|$ drops below the width of the PSF and the contributions to the image from the emitters begin to overlap, a phenomenon dubbed `Rayleigh's curse'~\cite{Tsang:2016-2}. This degradation can be understood by considering the spatial distribution of information content in the PSF. This is proportional to $\mathrm{dPSF}=\frac{\mathrm{d}I(x,y)}{ds}$, where $I(x,y)$ is the image intensity and $x$ and $y$ are the spatial coordinates~\cite{Fisher:1925,Chao:2016}. For conventional imaging, the maximum of $\mathrm{dPSF}$ is co-localised with the peak intensity of the image, where the optical shot noise is largest (Fig. \ref{fig:theory}A (left)), degrading the sensitivity of separation estimates. The split-Gaussian PSF overcomes this by modifying the PSF to localise the information in regions of low intensity (Fig. \ref{fig:theory}A (right)), similar to strategies employed for SPADE~\cite{Tsang:16,Rehacek:2017,Tsang:2017,Santamaria:2023}.

Previous SPADE techniques have been limited to one-dimensional imaging due to the use of one-dimensional spatial mode demultiplexing or PSF reshaping effective only along a single axis~\cite{Paur:18}. Our SDM protocol works in two-dimensions, which is important since practical super-resolution imaging should account for the typically unknown orientation of the emitter pairs being imaged. To validate this theoretically, we calculate the shot noise limited Fisher information as a function of both separation and orientation (Supplementary Text). Fig. \ref{fig:theory}B compares the Fisher information of SDM (red curve) and conventional microscopy (blue curve), along with their per-photon measurement resolution. Critically, the model predicts that SDM outperforms conventional microscopy in the useful far sub-diffracted regime. Specifically, the transition occurs at emitter separations $s$ below $0.45\sigma\approx90$~nm in visible light microscopy, independent of unstructured additive noise and regardless of emitter orientation (see Supplementary Text). A key benefit is that the split-Gaussian Fisher information decreases linearly as separation approaches zero due to the PSF's isolated zeros, compared to conventional microscopy where it decreases quadratically~\cite{Paur:18,Paur:2019}. This is one of the characteristic features of SPADE -- that the Fisher information degrades less slowly with decreasing separation than for conventional microscopy, with the advantage asymptotically approaching infinity as the separation approaches zero (Fisher information ratio, Fig. \ref{fig:theory}B (inset)). We also find that the Fisher information for SDM is highly robust to emitter orientation, varying by a maximum of 17$\%$ and an average of 5$\%$ across the full range of separations simulated (shaded band around red line, Fig. \ref{fig:theory}B).

\subsection*{Experimental Implementation}
Practical super-resolution imaging typically requires high-NA objectives to maximise resolution, together with suppression of background noise from the specimen and detection apparatus~\cite{Fish:2022}. We therefore implement SDM within a commercial TIRF microscope (3i Marianas): see Fig. \ref{fig:schematic}. TIRF microscopes excite fluorescence using the evanescent field from excitation light which is totally internally reflected from the coverslip surface, thereby confining the fluorescence to a thin layer (100-200~nm) near the glass interface~\cite{Mattheyses:2010,Fish:2022}. Our microscope employs an oil-immersion objective with a numerical aperture of 1.46 and a theoretical diffraction-limited resolution of 217~nm at a 519~nm imaging wavelength, based on the commonly used Rayleigh criterion for defining resolution limits in optical microscopy~\cite{Inoue:2006}.

To demonstrate the performance of SDM on a biologically relevant problem, we chose to image the length of self-assembled DNA nanorulers. These nanorulers have a well-characterised length due to their deterministic number of base pairs, with a 0.34~nm distance between adjacent bases~\cite{Watson:1953,Wilkins:1953,Franklin:1953}. We use fluorescently labelled nanorulers fixed on microscope slides of 50, 120 and 180 nm lengths (GATTA-quant), with manufacturer-specified $\pm10$~nm uncertainty arising primarily from curvature of the DNA segment. Twenty Alexa Fluor 488 fluorophores are clustered at each end, constituting the two incoherent emitters. They are excited with illumination light at 488~nm and fluoresce at 519~nm. Bandpass filters and dichroic beam splitters are used to separate the fluorescent emissions from the illumination field.

The illumination beam is focused onto the sample using a TIRF objective with 100$\times$ magnification, which is also used to collect the emission signal. The microscope column has an overall magnification of 840$\times$. We reshape the PSF by employing a quadrant phase plate, previously employed for two-dimensional centroid tracking~\cite{Treps:2003}. This shifts the phase of two diagonal quadrants by $\pi$ relative to the other two quadrants, leading to a more complex split-Gaussian PSF which is not rotationally symmetric. As predicted theoretically, we find that the property of spatially shifting signal away from noise is robust against emitter orientation. The phase plate, which we manufacture by dicing and gluing a Union Optic zero-order half waveplate (see Materials and Methods), is placed in the back focal (Fourier) plane of the microscope to alter the PSF of the collected field, and can be removed to compare the performance with conventional microscopy. A planoconvex lens then projects the image (focal) plane of the objective onto an ultra-low noise, high sensitivity EMCCD camera (iXon Ultra 897) which has 95\% quantum efficiency and single photon sensitivity. With cooling to -80~$^\circ$C, 300$\times$ gain and electron multiplication enabled, we achieve very low total electronic noise counts of around 0.13~s$^{-1}$ per pixel. 

To suppress ambient noise, we performed experiments in a dimly lit room with curtains over the equipment and lens tubes surrounding the optics; bandpass filters were used (three along each of the illumination and emission paths); backs of mirrors were painted black; and a low nanoruler concentration was used. This last modification reduced background fluorescence from the slide caused by fluorescent dye diffusing into the buffer solution -- the largest source of ambient noise in our experiment. With these modification, we achieve total ambient noise counts of around 2.3~s$^{-1}$ per pixel. 

The total number of emitted photons per emitter is limited by rapid photobleaching of the fluorophores. This restricts data collection to short time intervals, with images (each $N=512\times512$ pixels; $\text{field-of-view (FOV)}=85$~$\mu$m$^2$) captured over 500~ms exposure times. This exposure time also optimised for the effect of sample drift (which occurs over seconds~\cite{Dai:2016}).

Further details on experimental implementation challenges and how we overcome them, and experimental details including waveplate fabrication and alignment tolerances, are provided in the Materials and Methods.

\subsection*{Data Analysis}
\begin{figure}
    \centering
    \includegraphics[width=90mm]{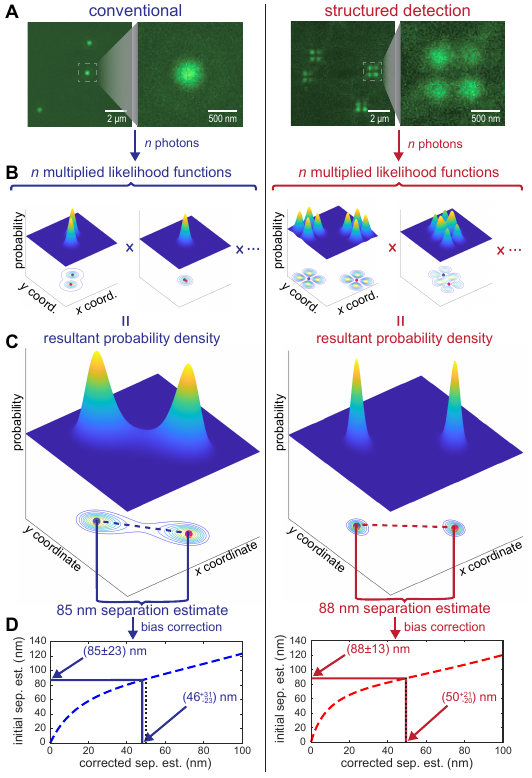}
    \caption{\textbf{Example (experimental) data and data analysis pipeline.} The left column is conventional microscopy, and the right SDM. (\textbf{A}) Example imaged PSFs. These photon intensity distributions include 32,851 and 43,419 photon detection events for the conventional and structured detection images, respectively. (\textbf{B}) The $n$ likelihood functions we construct from the imaged PSFs (the blue point is a photon's separation from the centroid, and the red point is the diagonally opposite separation; therefore, this represents the likelihood of the emitter separations from the centroid given a photon's separation). (\textbf{C}) The resultant probability densities for the emitter separations obtained by multiplying and normalising the $n$ likelihood functions. The maxima are taken to obtain the estimated emitter separations. These are substituted into bias correction functions, obtained via simulation under imaging conditions matching the actual image. (\textbf{D}) Corrected separation estimates. The black dashed line indicates the actual separation (50 nm).}
    \label{fig:data}
\end{figure}

To extract separation estimates from the acquired images we employ Bayesian analysis. Assuming the data is well described by a given model, Bayesian inference allows the posterior (post-measurements) joint probability distribution of the model parameters to be determined from noisy data. This can be used to compute the most probable estimate of a parameter given the data~\cite{Diaconis:1986,Toussaint:2011}. 

Our data analysis pipeline is detailed in Fig. \ref{fig:data}. Fig.~\ref{fig:data}A shows typical images obtained with conventional microscopy and SDM. Multiple emitter pairs are often visible within a single image, allowing parallel measurements. For data analysis, we crop the images to a smaller FOV of 7.4~$\mu$m$^2$ ($512\times512$ pixels) around each emitter pair (see Fig. \ref{fig:data}A). Each measurement collects tens of thousands of photons. The zoomed-in panels around individual emitter pairs reveal the expected PSFs, roughly Gaussian for conventional microscopy and four-lobed for SDM. In both cases, the contrast is relatively high, with an average of roughly 2.39 and 2.49 noise photons per pixel, and average peak pixel photon counts of around 38 and 24 respectively for conventional and SDM. While structured detection has lower contrast (ratio of peak signal photon number to noise photon number), we will see that the improved distribution of information relative to noise leads to a net increase in accuracy when estimating the separation. Experimentally, we find the diffraction limit is approximately 244~nm, a factor of 1.1$\times$ the theoretical value.

To model the images collected from the microscope, we assume random photon arrivals weighted spatially by the microscope PSF, and include uniform white noise from background photons and electronic dark noise. Each measurement is a photon detection event at a particular pixel coordinate. The DNA nanoruler emission is treated as incoherent, and we assume that the emitters are of equal brightness which is a good approximation for our experiments (see Materials and Methods). In total the model has seven parameters:  the $x$ and $y$ positions of each emitter, the $x$ and $y$ PSF widths and the magnitude of white noise relative to signal. The posterior distribution has dimensions equal to the number of parameters, while each measurement has a total number of possible outcomes equal to the number of pixels in the cropped image. Altogether, this makes direct Bayesian inference intractable, requiring storage of $N^7=3\times10^{30}$ elements.

As outlined in the Supplementary Text, to simplify the estimation problem we directly determine the mean positions of the emitters, the $x$ and $y$ PSF widths and the white noise level. The mean positions can be accurately found from the centroid of the PSF, and the white noise from the average noise level near the corners of the image away from the location of the DNA nanoruler emission. These parameter estimates do not suffer from Rayleigh's curse~\cite{Tsang:16}. As such, even a non-optimal estimate reduces parameter uncertainties below the separation uncertainty and does not significantly degrade the performance of SDM. Together, this reduces the problem to a tractable two parameters: the separations in each of the $x$ and $y$ directions $\{s_x,s_y\}$ from the centroid (see Supplementary Text), requiring storage of only $N^2=5\times10^8$ elements. We assume no prior information about the emitter separation.

\begin{figure*}
    \centering
    \begin{minipage}{105mm}
    \vspace{0pt}
    \centering
    \includegraphics[width=\linewidth]{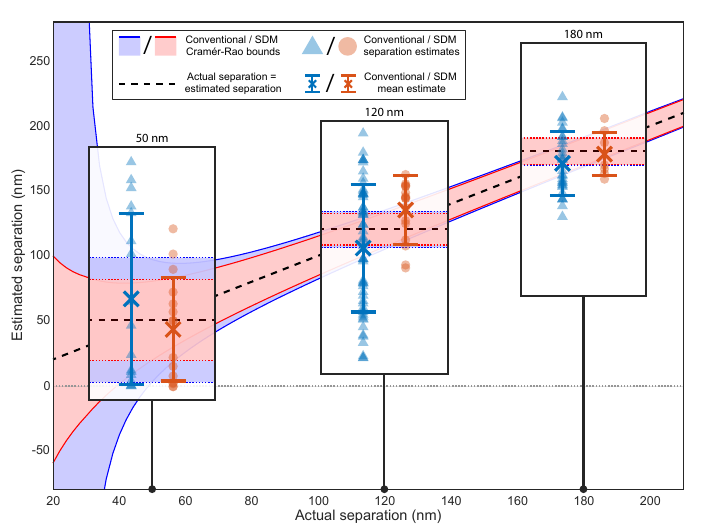}
    \end{minipage}
    \hfill
    \begin{minipage}{73mm}
    \vspace{0pt}
    \caption{\textbf{SDM and conventional microscopy results.} In this figure, blue is conventional microscopy and red is SDM. The outline of the coloured bands represent the Cram\'er-Rao bounds (CRBs) for the separation estimates from each method, with their width the theoretical resolution limit for any unbiased estimator. The SDM CRB is entirely subsumed within the conventional microscopy CRB under our imaging conditions. At each imaged separation, an overlay shows experimental data with identical vertical scaling to the figure. Error-bars represent the resolutions, and crosses are the means of each method. Individual separation estimates are represented with triangles for conventional microscopy (17, 58 and 31 estimates at 50, 120 and 180~nm length rulers, respectively), or circles for SDM (likewise, 15, 17 and 7 estimates). The dashed black line is where the actual separation equals the estimated separation. The $x$-axis maximum is approximately the ideal diffraction limit. A grey dashed line crosses the vertical axis at $y = 0$, and shows that with an ideal estimator, the two emitters cannot be reliably resolved below 49~nm (conventional) or 39~nm (SDM).}
    \label{fig:results}
    \end{minipage}
\end{figure*}

As is common in high resolution imaging, occasionally the observed PSFs are deformed, exhibiting aberrations or imperfect focusing. We manually exclude these images (see Materials and Methods). For each remaining image, a nonlinear least squares regression is used to determine the centroid, width and noise level. Under the assumption that the emitter separation is far sub-diffraction limited, this provides accurate model PSFs for each emitter. This allows us to determine the probability density of detecting a photon at a given position conditional on the separation, $P(x,y|s_x,s_y)$. Bayes' rule then provides the conditional probability density for the emitter separation given a photon detection event at location $\{x,y\}$, $P(s_x,s_y|x,y) \propto P(x,y|s_x,s_y)$. This corresponds to the likelihood function for $\{s_x,s_y\}$ assuming a uniform prior. The resultant (posterior) probability density is obtained by taking the product of the set of individual likelihood functions for each photon detection event and properly normalising (i.e., Bayesian inference). 

The distance between the two maxima in the resultant probability densities, shown in Fig. \ref{fig:data}C, provide an initial estimate for the emitter separation. These initial estimates are biased: the DNA nanoruler has a specified separation of 50 nm, whereas Bayesian inference predicts 85~nm (conventional) and 88~nm (SDM). Nonetheless, SDM provides around 1.8$\times$ better precision, consistent with theory (see the inset of Fig.~\ref{fig:theory}B, $x = 0.25\sigma$).

The bias is expected~\cite{Embacher:2022,Mortensen:2015}. To correct it we apply a correction function computed via simulation (Fig. \ref{fig:data}D, Supplementary Text). In the example of Fig. \ref{fig:data}, this results in separation estimates of $46^{+31}_{-23}$~nm and $50^{+21}_{-20}$~nm for conventional and structured detection, respectively, where the uncertainties are calculated by taking the full-width-half-maxima of the resultant probability densities and propagating this error through the bias correction functions. Both estimates are statistically consistent with the supplier-specified value of (50$\pm$10)~nm, validating our method, with SDM showing a $\sim32$\% reduction in uncertainty over conventional microscopy. We are also able to estimate the emitter orientation angle $\theta$ via the coordinate transformation $\{s_x,s_y\}\rightarrow\{s,\theta\}$. Projecting the probability densities from Fig. \ref{fig:data}C onto the $\theta$ axis provides $(-0.89\pm0.22)$~rad (conventional) and $(-0.88\pm0.17)$~rad (SDM), with SDM providing about 29$\%$ reduced uncertainty. Since the emitter pair angle is not known \textit{a priori}, the ground truth orientation is not known for these estimates and we could not validate these values beyond their self-consistency.  

\subsection*{Achieved Resolution}
To statistically validate the improvement that SDM provides over conventional microscopy, we made repeated measurements over an ensemble of emitter pairs. We define the root-mean-square-error (RMSE) in the separation estimates as the resolution. This provides a conservative upper-bound, since it includes the physical variation in the nanoruler length. Fig. \ref{fig:results} shows the resulting separation estimates, together with the means (crosses) and resolutions (error-bars) of each nanoruler length (50, 120 and 180~nm) for both conventional (blue) and structured detection (red). The means for both methods are statistically consistent with the specified lengths (dashed black lines), and as expected, the resolution degrades as separation decreases. SDM achieves resolutions of 39, 28 and 18~nm for 50, 120 and 180~nm length nanorulers, respectively, compared to 64, 53 and 27~nm for conventional microscopy. For 50~nm nanorulers, The resolution of SDM is 6.3$\times$ beneath our measured diffraction limit and 5.6$\times$ beneath the theoretical diffraction limit. Comparing the resolutions, we find that SDM outperforms conventional microscopy at all tested separations, with resolutions a factor of 1.6$\times$, 1.9$\times$ and 1.5$\times$ smaller for 50~nm, 120~nm and 180~nm rulers, respectively. 

The shaded regions in Fig. \ref{fig:results} (i.e., the CRBs) show the theoretical resolution limits, which are smaller than the resolutions achieved for both conventional and structured detection. This is expected since the RMSE provides a conservative estimate of resolution. At 50~nm separations, for example, the discrepancy between the experimental resolution of SDM (39~nm) and the Cram\'er-Rao bound (31~nm) is well explained by the specified $\pm$10~nm uncertainty in nanoruler length, as well as any residual bias or nonidealities in estimation like non-uniformity of background noise, EMCCD pixelation artifacts such as aliasing, the finite FOV when imaging, and deviations between the actual PSF and our models. At the smallest imaged lengths, even though SDM's resolution does not reach its theoretical limit, it still outperforms the resolution limit of conventional microscopy by a factor of $1.2\times$. 

\section*{Discussion}
Our work shows that far-sub-wavelength super-resolution imaging is possible by structuring the spatial mode of the detected field, and demonstrates this with biologically relevant samples. Integration within a state-of-the-art TIRF microscope, including the use of camera-based imaging and a phase plate rather than a spatial mode sorter, opens the door to practical applications in the life sciences. One limitation of our current experiments is their slow data processing, requiring approximately one hour of analysis on a standard desktop computer per image. Real-time data processing should be possible in future, since it has been implemented for Bayesian analysis of more complex datasets than ours~\cite{Kornak:2026,Henneking:2025,Cossio:2017}.

Our implementation of SDM achieves better than 40~nm resolution for all tested emitter separations with a half-second acquisition time. This resolution is primarily limited by background counts, and could be substantially improved by reducing them. For instance, our modeling predicts that a factor of ten reduction would allow 16~nm resolution for 50~nm separations (see Supplementary Text). The resolution could also be improved by collecting more photons, either by increasing the illumination intensity or acquisition time. We found that photobleaching occurred over 3~s, which indicates that a factor of six more photons could be collected. Combined with a factor of ten reduction in background, this could allow 5~nm resolution at 50~nm separations (see Supplementary Text). Further improvements may also be possible by optimising the spatial mode of the detected field to better separate the information-containing part of the field from the optical shot noise.

The resolution we demonstrate is already approaching that of many state-of-the-art super-resolution techniques. For instance, PALM, STORM and STED typically report lateral resolutions between 10-70~nm in ideal imaging conditions~\cite{Dempsey:2011,Khater:20,Lelek:21,Wu:2021}. SDM is therefore comparable to many state-of-the-art technologies, though not the more elaborate systems such as MINFLUX~\cite{Sahl:24}. Structured Illumination Microscopy (SIM) operates with standard fluorophores, but without photoswitching~\cite{Rego:2012} its resolution is capped at $\sim100$~nm~\cite{Ortkrass:23}.

Beyond resolution, most super-resolution techniques require optical illumination and many, such as STED or Ground State Depletion, require intense secondary illumination fields that increase the photon dose on specimens~\cite{Hell:94,Hell:95}. By contrast, SDM functions at low photon doses. It could even be applied to image intrinsically emitting systems such as bioluminescent and chemiluminescent reporters~\cite{Yeh:2019}, where external illumination is absent. These reporters encode functional biochemical activity at sub-wavelength length scales, but require fundamentally different approaches to break the diffraction barrier and obtain super-resolved images~\cite{Morales:2022,Siegel:2021}. SDM represents such an approach.

A key limitation of SDM is that it is not yet a general imaging approach -- rather, it can only determine the separation of pairs of emitters. This means it can currently only be applied to a narrow class of applications. Extending SDM to complex emitter distributions will require more elaborate systems, with some of the necessary strategies already proven in other implementations of SPADE~\cite{Matlin:2022}. Such an approach would require a sequence of measurements, which could be achieved by replacing the passive phase plate with an adaptive optical phase array. This has been shown to improve SIM~\cite{Debarre:2008} and other super-resolution microscopy techniques~\cite{Wang:2021}, and has been predicted to enable SPADE for complex emitter patterns~\cite{Titov:2025}. 


\bibliographystyle{ieeetr}
\bibliography{references} 


\section*{Acknowledgments}
Imaging was performed at the Queensland Brain Institute's Advanced Microscopy Facility using the 3i Marianas TIRF/FRET microscope.

K.C.R. acknowledges valuable discussions with Sam C. Scholten on preparation and presentation of figures.   


\newpage


\widetext
\appendix

\renewcommand{\thefigure}{S\arabic{figure}}
\renewcommand{\thetable}{S\arabic{table}}
\renewcommand{\theequation}{S\arabic{equation}}
\renewcommand{\thepage}{S\arabic{page}}
\setcounter{figure}{0}
\setcounter{table}{0}
\setcounter{equation}{0}
\setcounter{page}{1} 


\begin{center}
\section*{Supplementary Materials for \texorpdfstring{\\}{ } \papertitle}


Larnii~Booth$^{\ast}$,
Kyle~Clunies-Ross$^\ast$,
Rumelo~Amor,
Nicolas~Mauranyapin,\\
Zixin~Huang,
Michael~A.~Taylor,
Warwick~P.~Bowen$^\dagger$\\
\small$^\dagger$Corresponding author. Email: w.bowen@uq.edu.au\\
\small$^\ast$These authors contributed equally to this work.
\end{center}


\subsection*{Materials and Methods}
\subsubsection*{Quadrant waveplate fabrication and alignment}
We create our quadrant waveplates by using a Union Optic zero order half waveplate (achromatic between 450 and 650 nm) composed of two glued layers of different birefringent materials. The top layer has a phase shift of $(n+1)\lambda/2$ along a vertically-oriented fast axis, with the bottom layer having an $n\lambda/2$ phase shift along that same axis. This produces a total phase shift along the vertical axis of $\lambda/2$.  

\begin{figure}[t]
    \centering
    \includegraphics[width=90mm]{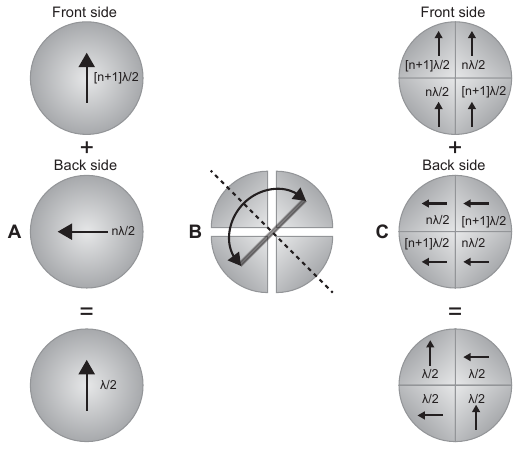}
    \caption{\textbf{Quadrant waveplate fabrication.} (\textbf{A}) A cemented zero order waveplate with two sides produces a total phase retardation of half a wavelength with a fast axis pointing vertically on one side and horizontally on the other side. (\textbf{B}) The waveplate is cut into four equal pieces, and the top-right and bottom-left pieces are flipped over a 45$^\circ$ axis and exchanged. (\textbf{C}) The result is the rearranged pieces have a phase shift that corresponds to a fast axis oriented vertically on the top-left and bottom-right quadrants, and horizontally on the top-right and bottom-left.}
    \label{fig:quadrantManufacture}
\end{figure}

We cut this waveplate into quadrants using an ADT7100 dicing saw with a resin blade for dicing glass, with the cut aligned parallel to the fast axis. We then swap the top-right and bottom-left quadrants by flipping them across the 45$^\circ$ axis joining the top-left and bottom-right quadrants (see Fig. \ref{fig:quadrantManufacture}). This gives the top-right and bottom-left quadrants a phase shift of $\phi(x)=1$ for these quadrants. This gives the desired structure for the quadrant waveplate in the Fourier plane: a $+1$ phase shift in the top-right and bottom-left quadrants, and a $-1$ phase shift in the top-left and bottom-right quadrants.

These pieces are glued to a beveled 2mm thick aluminium ring, with application along only the outer edge of the pieces to prevent glue running onto the waveplate's surface. We used Norland Optical Adhesive NOA 86 glue, a UV curing glue with high viscosity to prevent spreading over the waveplate surface during the curing process. 

To facilitate alignment, the waveplate was mounted on a 6-axis 1-inch optical mount allowing adjustment of the angle and rotation of the waveplate. Alignment involved checking that the position within the Fourier plane was correct, and that the four lobes of the PSF were roughly equal in brightness, by performing curvefitting on images of optical nanobeads with diameters of 200 nm. As explained below in the `Overcoming Experimental Challenges' section, the alignment tolerance was deemed met if the dimmest lobe was at least 40$\%$ as bright as the brightest lobe (with many being about 66$\%$ as bright), as simulations showed negligible impact on separation estimates at these tolerances.   

Convolution of the SDM phase plate shape with a Gaussian (to characterise waveplate diffusive scattering) was also performed, and showed no significant performance improvement -- indicating a minimal effect of waveplate imperfections on the obtained resolutions. As such, no convolution was performed in our data analysis.

\subsubsection*{Overcoming experimental challenges in biologically relevant samples}

\begin{figure}
    \centering
    \includegraphics[width=90mm]{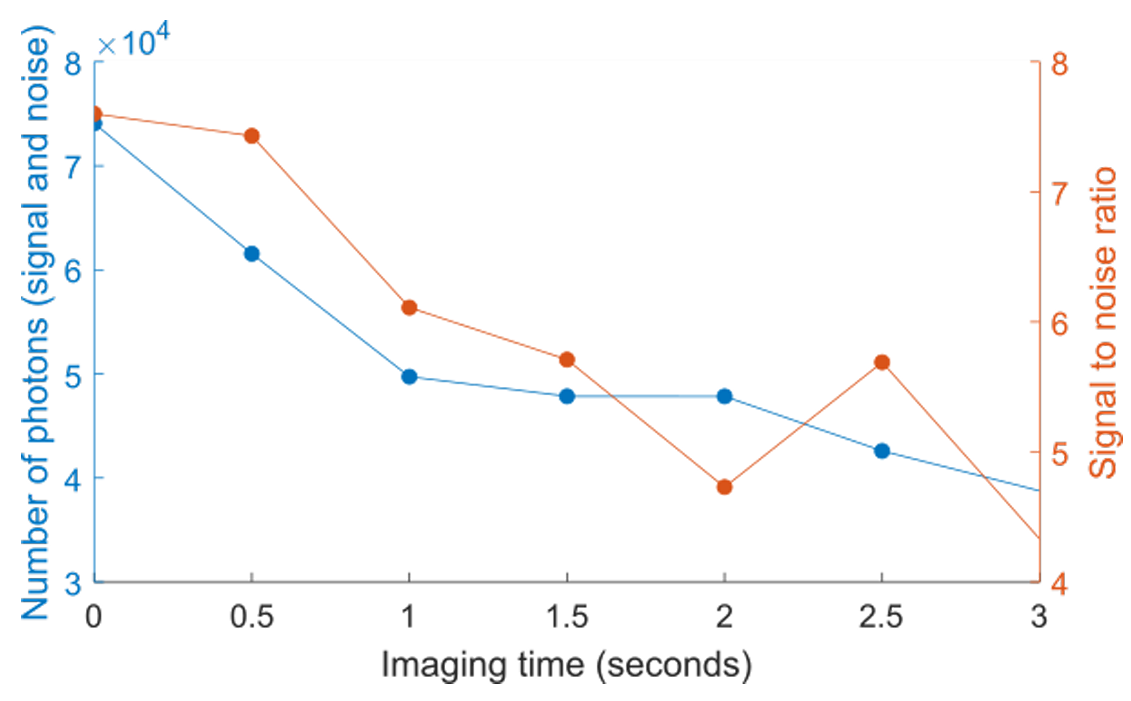}
    \caption{\textbf{Fluorescence decay of nanorulers.} Data from $s = 180$~nm nanorulers, imaged without a waveplate for 3~s. Left axis: total photons in the image. Right axis: signal to noise ratio, determined from a Gaussian curve fit. The measured intensity reduced by 50$\%$ over the imaging time, indicating that the fluorophore lifetime is 3~s.}
    \label{fig:photobleaching}
\end{figure}

Typical to biologically relevant samples, the fluorophores in our sample underwent rapid photobleaching, which limits data collection to short time intervals to avoid imaging background photons. TIRF microscopy alleviates this problem due to fluorophore excitation with the lower intensity evanescent field reducing the photobleaching rate~\cite{Hallworth:2012}. We also optimise the illumination intensity to balance dark-counts per second against read-out dark-counts in camera) extended our fluorophore lifetimes from the typical $<0.2$ s~\cite{Zhang:2012} to 3 s (based on a measured intensity reduction of 50\% after this time), as shown in Fig. \ref{fig:photobleaching}. Additionally, we used a rapid `scan-and-capture' method where the EMCCD camera was set to a video capture mode. The stage was then scanned in sections over 500 ms exposure times for each frame, ensuring that background photon capture past photobleaching was minimised. By using a longer exposure time, we also minimised any effects of fluorescence intermittency (`blinking') during continuous fluorophore excitation, which occurs on the order of milliseconds~\cite{Aitken:2008}. We note briefly here that because of the blinking duty cycle, this also effectively increases the required acquisition time to achieve a high SNR. 

To avoid sample drift (which typically limits resolutions for exposure times on the order of seconds \cite{Dai:2016}) and sample stage vibration (both of which can invalidate our model in parameter estimation and introduce aberrations), we used short ($<$ 1~s) exposure times. Nanoruler immobilisation with biotin and neutravidin on a BSA substrate also helped minimise these effects.

Optical aberrations~\cite{Bergmann:1999} are a key challenge when implementing SPADE in high resolution microscopes, since any aberrations change the PSF which can introduce inaccuracies in the model that significantly degrade separation estimates. To mitigate this, we calibrated the optical setup with 200 nm reference fluorescent nanobeads and `walked the beam path' (see~\cite{Kubitscheck:2013}). Furthermore, in data collection, we excluded nanorulers that laid far from the axis of the microscope, where aberrations are enhanced. We also excluded any images with deformed or blurry PSFs.

Equal emitter intensities were assumed, as numerical analyses indicated that non-equal intensities exhibited minimal effects ($<$ 10$\%$ error in the separation estimate) when the dimmer emitter was greater than 40$\%$ as bright as the other, which was always true during our experiments. 

Single emitter-pair images were also cropped to remove any effects from their long-tailed diffraction spikes, as these spikes introduce inaccuracies in the model degrading separation estimation performance. 

The effects of electronic noise \cite{Dussault:2004} were mitigated as follows: for dark current, we operated the camera below -70 $^\circ$C; for readout noise, we operated the camera at 300x electron multiplying gain to boost the signal prior to readout; and for clock induced charge, our EMCCD camera came equipped with compensatory electronics. Overall electronic noise counts under these conditions were as reported in the Main: 0.13~s$^{-1}$ per pixel.

To minimise ambient noise, we used multiple OD4 bandpass filters (three 488 $\pm$ 5 nm for the illumination beam and three 555 $\pm$ 44 nm for the emission beam), reducing laser added noise counts to 0.15~s$^{-1}$. Experiments were performed in a dimly lit room with curtains to separate the camera from the room and the room from the outside lab. The back of all mirrors in our setup were also painted black. Lens tubes were used to contain the optics. These changes reduced laboratory light ambient noise counts to 0.1~s$^{-1}$ per pixel. Finally, we made ample space between nanorulers to minimise ambient noise from nearby or diffused fluorescent dye. This was the largest source of noise in the experiment, and even with these modification, background noise counts originating from the slide were only reduced to between 1-3~s$^{-1}$ per pixel. On average, the total ambient noise count was as reported in the Main: around 2.3~s$^{-1}$ per pixel.

\subsubsection*{TIRF microscope modifications, pixel size calibration, and alignment}
\begin{figure}
    \centering
    \includegraphics[width=\textwidth]{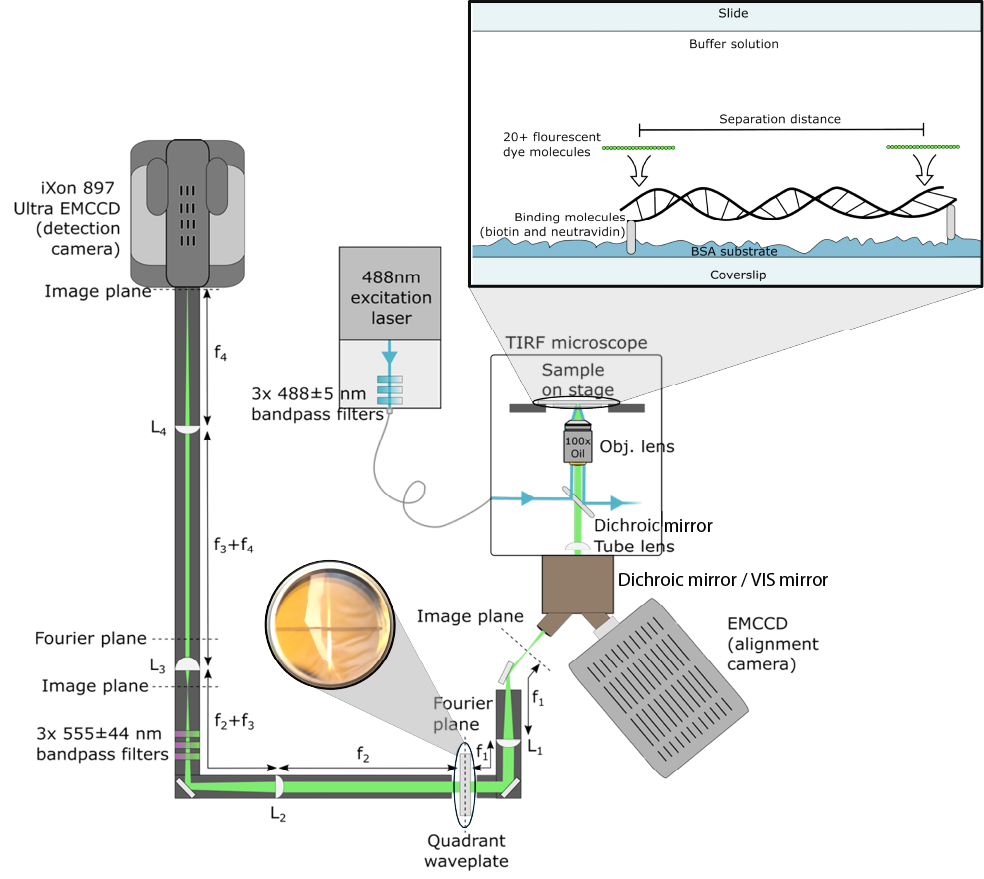}
    \caption{\textbf{Modifications to a TIRF Microscope to implement SDM.} Lenses $\mathrm{L}_1$ and $\mathrm{L}_2$ are used so that the quadrant waveplate can be inserted into the Fourier plane; lenses $\mathrm{L}_3$ and $\mathrm{L}_4$ are a simple telescope to magnify the resulting image for the detection camera. Bandpass filters are used to clean the excitation laser and reduce noisy photons in the emission light, and a long-pass dichroic mirror is used to remove the excitation laser light post-emission.}
    \label{fig:theory_supp}
\end{figure}

SDM was added into an existing 3i Marianas TIRF optical microscope.

To excite the fluorophores, a 488 nm excitation laser is used. The DNA nanorulers themselves emit light centred at a 519 nm wavelength. Three $(488 \pm 5)$ nm bandpass filters are used to clean the excitation laser light prior to excitation of the fluorophores, and three $(555 \pm 44)$ nm bandpass filters are used to remove as many non-signal photons from within the lens tube prior to detection. 

The DNA nanorulers light emission is collected using a 100x magnification oil immersion objective lens. A long-pass dichroic mirror reflects the excitation laser post-emission and transmits the fluorescence emission. We then send this light along two paths. One path ends in an EMCCD camera (the \textit{alignment camera}), while the other passes through a lens tube containing our passive optics to achieve SDM, which ends at another EMCCD camera (the \textit{detection camera}). Along this latter path, a plano-convex lens $\mathrm{L}_1$ is placed at its focal length in front of the image plane ($\mathrm{f}_1=100$ mm), collimating the beam to an approximately 2 cm width -- deliberately made large to reduce the impact of cracks and other edge defects on the waveplate. The waveplate was placed in the Fourier plane at a distance of $\mathrm{f}_2 = 200$ mm from another plano-convex lens $\mathrm{L}_2$. Lens $\mathrm{L}_2$ focuses the emission into the image plane. Two more lenses are then used to make a simple telescope, since otherwise the point sources are only a few pixels wide at the detection camera. Lens $\mathrm{L}_3$ ($\mathrm{f}_3 = 25$ mm) and $\mathrm{L}_4$ ($\mathrm{f}_4 = 200$ mm) magnify the PSF width by a factor of $200/25 = 8$. This light is then collected by an iXon 897 Ultra EMCCD camera. The alignment camera is a Photometrics Evolve 512 Delta EMCCD camera, and was used to find both the transverse and axial positions of the nanorulers. The microscope setup is detailed in Fig. \ref{fig:theory_supp}.

To calibrate the physical size per pixel, an image was taken with the alignment camera of an alignment grid with known dimensions. This gave a width of 151 nm per pixel in the image. For comparison, the same image was taken with the detection camera, which showed an 8.4x magnification factor (similar to the expected value of 8x). The width of each pixel in the detection camera was then computed as (151 nm)/(8.4x magnification) = 18 nm.

To correct for optical aberration, fluorescent nanobeads with 200 nm diameters were imaged without the waveplate and lens alignments were adjusted until aberrations fell within tolerance values. A curve fit was performed to determine the width of the Gaussian intensity profile in $x$ and $y$. A tolerance of up to 10 nm was deemed permissible due to confidence bounds in the fitting. 

Corrections for astigmatism were performed by walking the beam path until an out-of-focus PSF was not stretched in any direction. 

Noise sources were minimised according to the modifications mentioned in the `Overcoming Experimental Challenges' section.

Detected photon counts across the full image set ranged between approximately 20,000 to 160,000, with an average of approximately 67,000 for conventional microscopy and 59,000 for SDM per image. 

\clearpage
\newpage
\subsection*{Supplementary Text}

\subsubsection*{Data analysis pipeline}
\paragraph*{Likelihood functions and nonlinear curvefitting:}
We consider the two spatial modes of light detected in our experiments. 

The first spatial mode is the (ground-state) two-dimensional Gaussian mode, whose mean photon count at pixel $(i,j)$ is described by:
\begin{equation}
    \langle n\rangle_{i,j}=\frac{A}{\sigma^2}\left[\exp\left(-\frac{(x_{i,j}-\frac{s_x}{2})^2+(y_{i,j}-\frac{s_y}{2})^2}{\sigma^2}\right)+\exp\left(-\frac{(x_{i,j}+\frac{s_x}{2})^2+(y_{i,j}+\frac{s_y}{2})^2}{\sigma^2}\right)\right],
    \label{eqn:conventional}
\end{equation}
where we assume that the centroid is located at $(x_{i,j},y_{i,j})=(0,0)$ for simplicity, $(s_x,s_y)$ represent the separations of the emitters from the centroid ($+s$ for one emitter, $-s$ for the other), $\sigma$ is taken as the diffraction limit, and $A$ is a constant that accounts for the intensity of each emitter (assumed to be the same for both emitters).

For our first-order spatial mode that we shape with our quadrant waveplate, instead the mean photon count per pixel is given by:
\begin{multline}
    \langle n \rangle_{i,j} = \frac{A}{\sigma^6}\Bigg[\left(x_{i,j}-\frac{s_x}{2}\right)^2\left(y_{i,j}-\frac{s_y}{2}\right)^2\exp\left(-\frac{(x_{i,j}-\frac{s_x}{2})^2+(y_{i,j}-\frac{s_y}{2})^2}{3\sigma^2}\right)\\
    +\left(x_{i,j}+\frac{s_x}{2}\right)^2\left(y_{i,j}+\frac{s_y}{2}\right)^2\exp\left(-\frac{(x_{i,j}+\frac{s_x}{2})^2+(y_{i,j}+\frac{s_y}{2})^2}{3\sigma^2}\right)\Bigg].
    \label{eqn:SDM}
\end{multline}
Importantly, note that the spread of the photons shrink for each lobe of this spatial mode (hence the inverse proportionality to $\sigma^6$), and due to diffractive effects of the square EMCCD camera, we also have a smaller divisor in the exponential argument (inversely proportional to $3\sigma^2$).

The equations in these forms already incorporate shot-noise statistics, in that a simulation of photon arrivals using these equations will reproduce the expected Poissonian distribution of photon arrival counts. However, they do not account for the other sources of noise detailed in our main. Instead, we assume they collectively contribute a spatially uniform additive white Gaussian noise. To compute this noise, we take the corners of our images (where there are low photon counts from the emitters) and estimate the sample variance in photon counts as the noise $d$:
\begin{equation}
    d = \frac{\alpha}{N-1}\sum_{i}\sum_j (n_{i,j}-\langle n\rangle)^2,
\end{equation}
where $(i,j)$ runs over the pixels denoting the corners of each image, $\langle n \rangle$ is the mean pixel count over all corners, and $\alpha$ is an \textit{ad hoc} constant dependent on the imaged spatial mode that accounts for diffractive effects that cause some signal photons to bleed into the image corners (for the Gaussian mode, $\alpha = 1$, and for the split-Gaussian mode, $\alpha = 1.1$). 

To estimate the centroid, $A$ and $\sigma$ of the emitters, we take our images and pass them into MATLAB's inbuilt \texttt{fit} function, where instead of $x_{i,j}-\frac{s_x}{2}$ and $y_{i,j}-\frac{s_y}{2}$, we use $x_{i,j}-b_{x,1}$ and $y_{i,j}-b_{y,1}$ (and equivalently, $b_{x,2}$ and $b_{y,2}$ for $(+\frac{s_x}{2},+\frac{s_y}{2})$), where $b$ is a variable representing the location of the emitters. The functions we fit to are:
\begin{equation}
    \langle n \rangle_\text{fit} =  
    \begin{cases}
        \frac{A}{\sigma^2}\left[\exp\left(-\frac{(x-b_{x,1})^2+(y-b_{y,1})^2}{\sigma^2}\right)+\exp\left(-\frac{(x-b_{x,2})^2+(y-b_{y,2})^2}{\sigma^2}\right)\right]+d\quad\text{ if conv.}\\
        \frac{A}{\sigma^6}\Big[(x-b_{x,1})^2(y-b_{y,1})^2\exp\left(-\frac{(x-b_{x,1})^2+(y-b_{y,1})^2}{3\sigma^2}\right)\\
        \qquad+(x-b_{x,2})^2(y-b_{y,2})^2\exp\left(-\frac{(x-b_{x,2})^2+(y-b_{y,2})^2}{3\sigma^2}\right)\Big]+d\quad\text{ if SDM.}
    \end{cases}
    \label{eqn:fit}
\end{equation}
Because nonlinear curvefitting is prone to incorrectly producing parameters that are locally (though not globally) optimal, we run the \texttt{fit} function multiple times, shrinking the allowed locations for the emitter centres (we observed that $\sigma$ estimates were correlated to these values, so did not need to also adjust for $\sigma$). We then take the curve whose parameters give the largest $r^2$ value. From this, we compute the centroid coordinates as:
\begin{equation}
    c_x=\frac{b_{x,1}+b_{x,2}}{2},\,\,\,c_y=\frac{b_{y,1}+b_{y,2}}{2}.
\end{equation}
Accounting for this centroid, new photon coordinates are computed as $x'_{i,j}=x_{i,j}-c_x$ and $y'_{i,j}=y_{i,j}-c_y$. We substitute $x'_{i,j}$ and $y'_{i,j}$ for $x_{i,j}$ and $y_{i,j}$ into Equations \ref{eqn:conventional} (for conventional microscopy) and \ref{eqn:SDM} (for SDM) when using them during our Bayesian inference step. 

In order to compute the likelihood functions, we use the \textit{normalised} $\langle n \rangle$ equations. Since we are attempting to compute $(s_x,s_y)$, we prepare a 2D grid across the domain of the image; since we pass in $x'_{i,j}$ and $y'_{i,j}$, each value in this grid represents a possible estimate for $\frac{s}{2}$. The peaks of the resultant probability densities will approximate the maximum likelihood estimates for $\frac{s}{2}$, and the distance between these points will give us $s$. Due to aliasing effects, it is important to supersample the points in $s$ compared to the actual pixel spacing in the image (if, for example, the emitters were actually centred \textit{within} a pixel instead of at the exact boundaries of a pixel). We choose a value for the $s$ spacing that is computationally tractable, yet still reduces aliasing; in our experiment, we chose a spacing $1/18$th the physical pixel grid spacing. We then randomly sample the coordinates for every detected photon in the image to avoid any spurious zeros due to floating point errors. We pass in these coordinates for $x'_{i,j}$ and $y'_{i,j}$ alongside our $s$ grid. To further avoid any floating point errors, we then add the $\log$ of each $\langle n \rangle$ function (equivalent to multiplying each $\langle n \rangle$), and exponentiate the result to obtain our resultant probability densities.  

\paragraph*{Bayesian inference details:}
Given two random variables $A$ and $B$ with probability distributions $P(A)$ and $P(B)$, Bayes' theorem reads:
\begin{equation}
P(A|B) = \frac{P(B|A)P(A)}{P(B)}
\end{equation}
where $P(B|A)$ is the probability of $B$ given $A$, and hence $P(A|B)$ is the probability of $A$ given $B$. 

In our work, we assume there are 2 emitters in the region of interest. Then there are four parameters of interest: the 2D coordinates of emitter 1, $(x_1,y_1)$, and likewise for emitter 2, $(x_2,y_2)$. Our observations are photon positions, and not knowing the emitter for a particular detected photon, we assign coordinates $(X_1,Y_1)$ for photon detection coordinates with respect to emitter 1, and likewise $(X_2,Y_2)$ for emitter 2. Conditioned on a single detection event, we can denote the emitter coordinates $A = [x_1,y_1,x_2,y_2]$, and the photon arrival coordinates $B = [X_1,Y_1,X_2,Y_2]$.

It is computationally inefficient to apply Bayes' rule using our current definitions for $A$ and $B$ since this is a 4-dimensional parameter space. Consider instead a parameterisation with the centroid coordinates $(c_x,c_y) = \big(\frac{x_1+x_2}{2},\frac{y_1+y_2}{2}\big)$, and the separation between the emitters $(s_x,s_y)=(x_2-x_1,y_2-y_1)$. We can now rewrite $A$ as:
\begin{equation}
    A=\Big[c_x-\frac{s_x}{2},c_y-\frac{s_y}{2},c_x+\frac{s_x}{2},c_y+\frac{s_y}{2}\Big].
\end{equation}
This requires estimating the centroid first, something already obtained from our nonlinear least-squares fit. As mentioned in the Data Analysis section of our main, the estimation of the centroid -- unlike the separation -- does not suffer from Rayleigh's curse and so even a non-optimal estimate reduces the centroid uncertainty below the separation uncertainty and does not degrade the performance of SDM.

We then consider $B$ not as the coordinates with respect to each emitter, but the separations with respect to the centroid:
\begin{equation}
    B=\Big[\frac{s_x}{2},\frac{s_y}{2},-\frac{s_x}{2},-\frac{s_y}{2}\Big].
\end{equation}
The first and last two terms in $A$ and $B$ are now symmetric about the centroid and so the parameter space reduces to 2 dimensions. 

We now employ Bayesian inference by iteratively applying Bayes' rule for $n$ detection events. Begin by assuming a flat, uniform prior, representing that any emitter position is equally probable: $P_0(A)=\frac{1}{d^2}$, with $d$ the number of pixels along one dimension in a square grid, and the subscript the number of iterative applications of Bayes' rule (in general, we term this $i$). Bayesian inference sets the next prior to be the previous posterior, such that $P_{i}(A)=P_{i-1}(A|B)$. Then we have in general:
\begin{equation}
    P_{i}(A|B)=\frac{P_{i}(B|A)P_{i-1}(A|B)}{P_{i}(B)},
    \label{eqn:BayesRule}
\end{equation}
where, for clarity, $P_0(A|B)=P_0(A)$.

We can simplify since the posterior, being a probability density, must be normalised. This allows us to write $P(B) = \sum P_i(B|A)P_{i-1}(B|A)$. Hence, applying Eqn. \ref{eqn:BayesRule}, the resultant probability density is the product of normalised likelihood functions times priors over all observed photons $i=1$ to $n$. Since $P_0(A)$ is uniform, and since the likelihood functions (being probability densities themselves) are normalised, hence the posterior will just be the normalised product of likelihood functions. We then obtain the resultant posterior distribution:
\begin{equation}
    P_n(A|B)=\frac{\prod_{i=1}^{n}P_i(B|A)}{\sum \prod_{i=1}^{n}P_i(B|A)}.
    \label{BayesianInference}
\end{equation}

Initially, the analytical forms of $P(A|B)$ are the nonlinear fitting functions as described in the prior section of this Supplementary Text. Iterative application of Bayes' rule causes $P(A|B)$ to converge to the sum of two Gaussians, as shown in Fig. \ref{fig:data}c. We can now take the maxima of these resultant probability densities as the initial estimate of the separation.

\paragraph*{Bias correction functions:}
As noted in the Data Analysis section in our main, there is inherent biasing caused by our Bayesian inference method. However, for a given image with similar levels of noise and $\langle n \rangle_\text{fit}$ parameters, we assume this biasing is consistent and hence reversible via a bias correction function. 

Since separation estimates are biased away from the true separation for small $s$, and converge to the true separation for large $s$ (above the diffraction limit), this bias correction function needs to take $s$ values and convert them to smaller ones at small $s$, while leaving larger $s$ values unchanged. It therefore must transition from a linear regime to a non-linear regime below some $s$ value. Presumably, the point at which these regimes switch over, and the steepness of the non-linear region, should also depend on the noise and $\langle n \rangle_\text{fit}$ parameters for a given set of images. The function should also monotonically increase for $s > 0$. An \textit{ad hoc} implicit function which achieves these requirements at positive $s$ values would be $s_\text{corrected}$ divided by a modified sigmoid function:
\begin{equation}
    s = s_\text{corrected}\left({\frac{1+\beta}{1+\exp(-\gamma s_\text{corrected})}-\beta}\right)^{-1},
    \label{eqn:bias_fit}
\end{equation}
where $\beta$ and $\gamma$ are fitting parameters. This function is invertible for all $s_\text{corrected}>0$, however it has no general analytic closed-form expression. Nonetheless, an estimate of an inverted value is possible via numerical root-finding, and any numerical errors will still be smaller than the RMSE.

To obtain this function, simulations of photon arrivals are needed. These simulations are based on the average parameters from the $\langle n \rangle_\text{fit}$ of all imaged data, namely $A$, $\sigma$, $d$ and $n$ (the total number of photons in the image). We simulated $n$ random photon arrivals weighted by a normalised $\langle n \rangle_\text{fit}$ across different $s$ values (where we can assume here the centroid is located at the centre pixel of our camera). We then ran multiple iterations of our nonlinear curvefitting and Bayesian inference process (identically to our real data) at each of these separations, and from this simulated data we can obtain our $\beta$ and $\gamma$ parameters using a nonlinear least squares fit on Eqn. \ref{eqn:fit} since we know the true underlying $s_\text{corrected}$ values. This approach also allows us to compute the confidence bounds in Eqn. \ref{eqn:bias_fit}, however these contribute more than an order of magnitude smaller errors than the RMSE and hence negligibly contribute to the errors in our estimates. 

Two example bias correction functions, including the 95$\%$ confindence bounds on the $\beta$ and $\gamma$ parameters, are provided in Eqn. \ref{eqn:bias_fit_examples} below for the images of the 50 nm DNA nanorulers. 
\begin{equation}
    s = \begin{cases}
        s_\text{corrected}\left({\frac{1+(0.7111\pm0.0586)}{1+\exp(-(0.0174\pm0.0006)s_\text{corrected})}-(0.7111\pm0.0586)}\right)^{-1}\quad\text{ if conv.}\\
        s_\text{corrected}\left(\frac{1+(0.8140\pm0.1058)}{1+\exp(-(0.0236\pm0.0014)s_\text{corrected})}-(0.8140\pm 0.1058)\right)^{-1}\quad\text{ if SDM.}
        
    \end{cases}
    \label{eqn:bias_fit_examples}
\end{equation}

For reference, these are the bias correction functions included in Fig. \ref{fig:data}d.

Since there is no reliance upon knowledge of the actual $s$ values being imaged, therefore our bias correction functions do not feed in any assumed knowledge for $s$, and can be computed in parallel to the biased $s$ values from the imaged data.

\paragraph*{Mean experimental separation estimates and errors:}
For nanorulers of 50~nm length, SDM estimates a mean separation of $43 \pm 10$ nm compared to $66 \pm 15$ nm for conventional microscopy, where the errors are the standard error of the mean. At 120~nm length, the mean separations were $136\pm 6$~nm and $105\pm 7$~nm for SDM and conventional microscopy, respectively; while at 180~nm they were $178 \pm 7$~nm and $170 \pm 5$~nm.

\subsubsection*{Fisher information computation}
The Fisher information -- which gives the amount of information a dataset contains about a specific variable -- is related to the mean-squared-error via:
\begin{equation}
    \mathrm{FI}\leq\frac{1}{\mathrm{MSE}},
\end{equation}
with $\mathrm{FI}$ the Fisher information, and $\mathrm{MSE}$ the mean-squared-error. This bounds the smallest possible RMSE of any unbiased estimator according to:
\begin{equation}
    \mathrm{RMSE} \geq \mathrm{CRB} = \frac{1}{\sqrt{\mathrm{FI}}},
\end{equation}
with $\mathrm{CRB}$ the `Cram\'er-Rao bound' (the smallest RMSE achievable by an unbiased, optimal estimator).

The general expression for the Fisher information for a signal $\theta$ of an image with $n$ photons to determine value $x$ is:
\begin{equation}
    \mathrm{FI}_x(\theta) = n\mathbb{E}\left[\left(\frac{\partial}{\partial \theta}\log(f(x|\theta))\right)^2\right],
\end{equation}
where $\mathbb{E}$ is the expectation value of the function, and $f(x|\theta)$ is the probability distribution of the signal $\theta$ for observation $x$. 

In a two dimensional image, the discrete Fisher information for separation (which we use to account for image pixelation) can be rewritten as:
\begin{equation}
    \mathrm{FI}_{x,y}(s) = n\sum_{i,j} \left[\frac{\partial}{\partial s}(f(x_i,y_j|s)\right]^2 \frac{1}{f(x_i,y_j|s)},
\end{equation}
where the sum is over all pixels $(i,j)$, and $f(x_i,y_j|s)$ is the probability that a photon lands on a pixel located at coordinates $(x_i,y_j)$ given a specific separation. These probability densities are given by the normalised $\langle n \rangle_\text{fit}$ fits given in Eqn. \ref{eqn:fit}. We do not account for the non-orthogonal contributions to the Fisher information from other parameters (such as emitter brightness) to simplify the analysis here, though we note that a full analysis would further reduce the Fisher information by these contributions.

To compute the ideal (i.e., shot-noise limited) Fisher information, we simulate photon arrivals using Eqn. \ref{eqn:fit} for both conventional and structured detection with $d = 0$. As the conventional modeshape is rotationally symmetric, its Fisher information is independent of emitter orientation. To quantify the effect of the Fisher information for structured detection, whose modeshape is more complex owing to its asymmetry with respect to emitter orientation, we simulate photon arrivals for emitter separations at 4 different angles: 0, $\pi/12$, $\pi/6$ and $\pi/4$ radians (no larger angles were necessary owing to the quadrant symmetry of the modeshape). In Fig. \ref{fig:theory}(b), the mean value from these simulations is the thick red line, whereas the shaded outline represents the boundary of all four Fisher information curves when plotted simultaneously. We note here that the critical point in Fig. \ref{fig:theory}b was independent of $d$, though the per-photon resolution reduced with increasing $d$. 

To compute the Fisher information for our image data, we take the average values for $A$, $\sigma$, $n$ and $d$ from these images for each nanoruler separation and imaging methodology (conventional microscopy or SDM), as well as the physical and supersampled pixel size. We also account for the larger area in the domain of our probability densities by adjusting the photon number upwards to include the larger number of signal photons (this larger domain is necessary to confirm the results converge to expected values outside of super-resolved separations). 

To determine the $A$, $n$, $\sigma$ and $d$ values between our nanoruler lengths, we interpolate using a linear quadratic fit for each parameter from the three datapoints at 50, 120 and 180 nm between the values of $s = 30$ and $s = 200$ nm. Outside these lengths, we assume the values are constant (eg, the value for $d$ below 30 nm is the value for $d$ \textit{at} 30 nm according to our interpolated fit). This produces a kink due to the discontinuity in the derivative at these points, however the effect on the values in between these points is negligible. Since we only use values outside these lengths to confirm the behaviour converges to expected performance as $s \rightarrow 0$ or $s \rightarrow \infty$, this means the approach is acceptable for the purposes of our research.   

Since the spatial modes are symmetric for conventional microscopy, but asymmetric for SDM, along $x$ and $y$, for simplicity we only compute the derivative by changing just the $x$ separation for conventional microscopy; in principle, this would converge to the estimate for separation changes along any direction with a small enough pixel spacing in the conventional case. For SDM, as discussed above, we alter the separation angle in 15$^\circ$ ($\pi/12$ rad) increments up to 45$^\circ$ ($\pi/4$ rad), and take the minimum and maximum values of the FI over these angles as the envelope in Fig. \ref{fig:theory}. Finally, to numerically estimate the derivative, we use a first-order central finite difference method. 

\begin{figure}
    \centering
    \includegraphics[width=\textwidth]{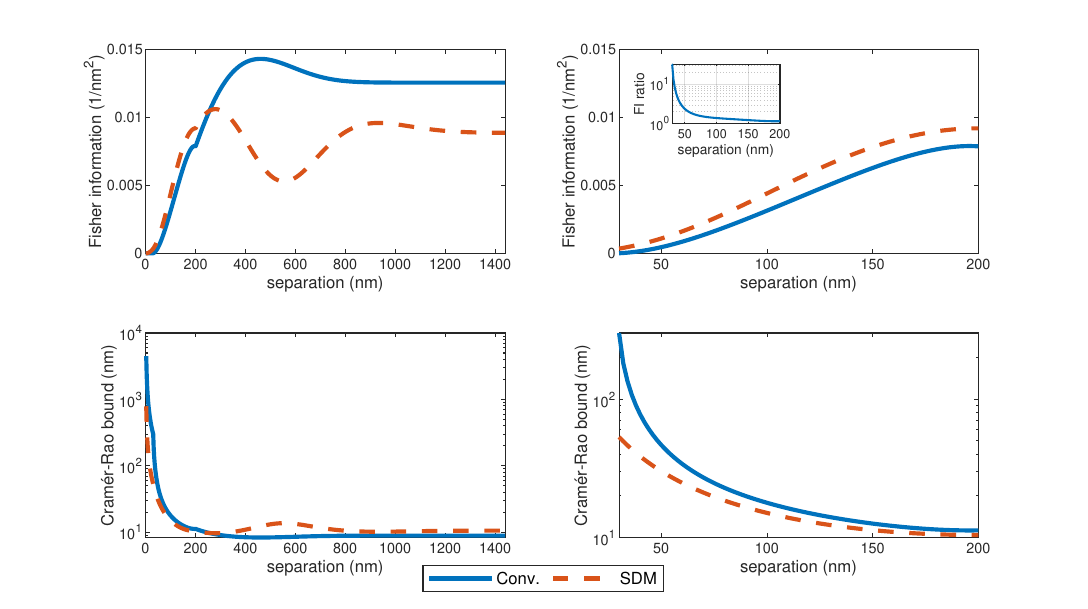}
    \caption{\textbf{Numerical results for the Fisher information and Cram\'er-Rao bounds.} Fitting parameters are taken from our images and applied over a wide separation range. The top row is Fisher information, and the bottom row is the Cram\'er-Rao bound. Leftmost plots are over the entire simulated separation range, and rightmost plots are a smaller range covering the far sub-diffracted region relevant to our results. The critical point shift and quadratic shape of the SDM FI as $s \rightarrow 0$ is caused by diffractive effects from the EMCCD camera geometry, different values for the fit parameters of $\langle n \rangle_\text{fit}$, and the presence of random noise. Despite this, the CRB for SDM is still less than for conventional microscopy at super-resolved separations. The kinks at $s = 30$ nm and $s = 200$ nm are caused by discontinuities in the derivatives for the $\langle n \rangle_\text{fit}$ parameters at these separations, and have a negligible effect on the FI and CRB between these separations. The inset in the top-right plot shows the FI ratio of conventional microscopy to SDM, which diverges as $s \rightarrow 0$.}
    \label{fig:FI_supp}
\end{figure}

Numerical results for the Fisher information and the Cram\'er-Rao bound based on our image data from $s = 2$ nm to $s = 1440$ nm are provided in Fig. \ref{fig:FI_supp}; for completeness, these are just the values with the separation only along the $x$-axis. As can be seen, due to both diffractive effects from the EMCCD geometry (causing a difference in lobe width for SDM), different parameter values between conventional microscopy and SDM and the presence of random noise, the resolution limit where SDM `beats' conventional microscopy is moved further out to approximately 254 nm. These effects also cause the Fisher information of SDM to converge quadratically at small separations, however as can be seen from the plot of the Cram\'er-Rao bound, this still produces a smaller RMSE for our super-resolved images.

From these results, we can also model the effect of reduced background noise, increased emission / acquisition times, or both simultaneously. We adopt a simplistic approach wherein increasing acquisition times simply increases the values for $A$, and decreasing background noise reduces the values for $d$, from Equation \ref{eqn:fit}. For example, by increasing acquisition time by a factor of 6x, and decreasing background by a factor of 10x, this results in $A_\text{new}=6A$ and $d_\text{new}=d/10$. The CRB value at 50~nm with these parameters is reported as our predicted resolution limit in the Discussion of the main text.

\end{document}